\documentclass[%
 prapplied,
 amsmath,amssymb,
 superscriptaddress,
 reprint,%
floatfix, 
]{revtex4-2}
\usepackage{graphicx}
\usepackage[utf8]{inputenc}
\usepackage[T1]{fontenc}
\usepackage{mathptmx}
\usepackage{dcolumn}
\usepackage{etoolbox}
\usepackage[caption=false]{subfig}
\usepackage{makecell}
\usepackage{bm}
\usepackage{multirow}

\newcommand{\SiN}[0]{Si$_3$N$_4$}

\newcommand{\nint}[0]{$\overline{n}$}
\newcommand{\inverseQint}[0]{$Q_{\text{i}}^{-1}$}

\newcommand{\inverseQext}[0]{$Q_{\text{e}}^{-1}$}

\newcommand{\pSiN}[0]{$F_\text{SiN}$}

\newcommand{\dif}{\mathop{}\!\mathrm{d}}
\newcommand*\mean[1]{\bar{#1}}
\DeclareMathOperator{\sech}{sech}

\begin{document}
\title{Annealing reduces \SiN{} microwave-frequency dielectric loss in superconducting resonators}

\author{S. Mittal}
\altaffiliation{These authors contributed equally to this work.\\Direct correspondence to sarang.mittal@colorado.edu}
\affiliation{JILA, National Institute of Standards and Technology and the University of Colorado, Boulder, Colorado, 80309, USA}
\affiliation{Department of Physics, University of Colorado, Boulder, Colorado, 80309, USA}

\author{K. Adachi}
\altaffiliation{These authors contributed equally to this work.\\Direct correspondence to sarang.mittal@colorado.edu}
\affiliation{JILA, National Institute of Standards and Technology and the University of Colorado, Boulder, Colorado, 80309, USA}
\affiliation{Department of Physics, University of Colorado, Boulder, Colorado, 80309, USA}

\author{N. E. Frattini}
\affiliation{JILA, National Institute of Standards and Technology and the University of Colorado, Boulder, Colorado, 80309, USA}
\affiliation{Department of Physics, University of Colorado, Boulder, Colorado, 80309, USA}

\author{M. D. Urmey}
\affiliation{JILA, National Institute of Standards and Technology and the University of Colorado, Boulder, Colorado, 80309, USA}
\affiliation{Department of Physics, University of Colorado, Boulder, Colorado, 80309, USA}

\author{S-X. Lin}
\affiliation{JILA, National Institute of Standards and Technology and the University of Colorado, Boulder, Colorado, 80309, USA}
\affiliation{Department of Physics, University of Colorado, Boulder, Colorado, 80309, USA}

\author{A. L. Emser}
\affiliation{JILA, National Institute of Standards and Technology and the University of Colorado, Boulder, Colorado, 80309, USA}
\affiliation{Department of Physics, University of Colorado, Boulder, Colorado, 80309, USA}

\author{C. Metzger}
\affiliation{JILA, National Institute of Standards and Technology and the University of Colorado, Boulder, Colorado, 80309, USA}
\affiliation{Department of Physics, University of Colorado, Boulder, Colorado, 80309, USA}

\author{L. Talamo}
\affiliation{JILA, National Institute of Standards and Technology and the University of Colorado, Boulder, Colorado, 80309, USA}
\affiliation{Department of Physics, University of Colorado, Boulder, Colorado, 80309, USA}

\author{S. Dickson}
\affiliation{JILA, National Institute of Standards and Technology and the University of Colorado, Boulder, Colorado, 80309, USA}
\affiliation{Department of Physics, University of Colorado, Boulder, Colorado, 80309, USA}

\author{D. Carlson}
\affiliation{Octave Photonics, LLC., Boulder, Colorado, 80305, USA}
\affiliation{Department of Physics, University of Colorado, Boulder, Colorado, 80309, USA}

\author{S. B. Papp}
\affiliation{Time and Frequency Division, National Institute of Standards and Technology, Boulder, Colorado, 80305, USA}
\affiliation{Department of Physics, University of Colorado, Boulder, Colorado, 80309, USA}

\author{C. A. Regal}
\affiliation{JILA, National Institute of Standards and Technology and the University of Colorado, Boulder, Colorado, 80309, USA}
\affiliation{Department of Physics, University of Colorado, Boulder, Colorado, 80309, USA}

\author{K. W. Lehnert}
\affiliation{JILA, National Institute of Standards and Technology and the University of Colorado, Boulder, Colorado, 80309, USA}
\affiliation{Department of Physics, University of Colorado, Boulder, Colorado, 80309, USA}

\begin{abstract}
    \noindent The dielectric loss of silicon nitride (\SiN{}) limits the performance of microwave-frequency devices that rely on this material for sensing, signal processing, and quantum communication. Using superconducting resonant circuits, we measure the cryogenic loss tangent of either as-deposited or high-temperature annealed stoichiometric \SiN{} as a function of drive strength and temperature. The internal loss behavior of the electrical resonators is largely consistent with the standard tunneling model of two-level systems (TLS), including damping caused by resonant energy exchange with TLS and by the relaxation of non-resonant TLS. We further supplement the TLS model with a self-heating effect to explain an increase in the loss observed in as-deposited films at large drive powers.
    Critically, we demonstrate that annealing remedies this anomalous power-induced loss, reduces the relaxation-type damping by more than two orders of magnitude, and reduces the resonant-type damping by a factor of three. 
    Employing infrared absorption spectroscopy, we find that annealing reduces the concentration of hydrogen in the \SiN{}, suggesting that hydrogen impurities cause substantial dissipation.
\end{abstract} 

\maketitle

\section{Introduction}
Silicon nitride \SiN{} is a crucial material for a range of applications at microwave frequencies. Radio frequency micro-electromechanical systems leverage the unique mechanical properties of the dielectric to realize sensors, actuators, and switches \cite{kaloyeros_reviewsilicon_2020, kurmendra_review_2021}.
Suspended \SiN{} membranes help achieve excellent thermal isolation of transition-edge sensors in bolometer and calorimeter applications \cite{irwin_transition-edge_2005, ullom_review_2015}. These membranes are also used to create extraordinarily high quality factor mechanical resonators \cite{thompson_strong_2008, yuan_silicon_2015, ghadimi_elastic_2018, reetz_analysis_2019} and are used in experiments that couple electrical and mechanical resonators with the goal of achieving quantum control of motion \cite{naik_cooling_2006}. These electromechanical experiments have realized ground-state cooling of macroscopic resonators \cite{fink_quantum_2016, seis_ground_2022}, demonstrated efficient microwave frequency conversion \cite{fink_efficient_2020}, and shown promise for quantum-enhanced sensing \cite{bothner_cavity_2020}.
Finally, there are efforts to use these mechanical resonators for quantum microwave-to-optical frequency conversion \cite{andrews_bidirectional_2014, higginbotham_harnessing_2018, planz_towards_2022}, leveraging the ability of \SiN{} mechanical resonators to interact with both electrical and optical resonators. 

The electromechanical results in particular share a common feature: a strong microwave drive tone is required to maximize the parametric coupling of the microwave-frequency resonant circuit and mechanical resonator. The drive tone populates the circuit with photons and enhances the electromechanical interaction rate, but at the same time may induce microwave loss, a resonance frequency shift, or other nonlinear effects.
In fact, recent work using \SiN{} membranes for microwave-to-optical frequency conversion was limited by excess power-induced loss in the electrical circuit \cite{delaney_superconducting-qubit_2022, brubaker_optomechanical_2022}. Understanding and eliminating the contribution to that microwave loss from the \SiN{} dielectric is important for future advancements with membrane electromechanical systems.

Hydrogen impurities are a natural suspect for this loss, as they are known to be present in \SiN{} films \cite{chow_hydrogen_1982, parsons_low_1991} and contribute to low-power microwave dielectric absorption \cite{paik_reducing_2010}.
Meanwhile, nanophotonics applications commonly perform a high temperature, post-deposition thermal anneal to remove residual hydrogen impurities from the precursor gases used in the deposition of nominally stoichiometric \SiN{} films \cite{luke_overcoming_2013, ji_methods_2021}. Together, these observations imply that annealing could improve the dielectric loss of \SiN{} by eliminating hydrogen impurities.

In this work, we investigate the effect of annealing on the power- and temperature-dependent microwave dielectric loss of \SiN{}. Specifically, we fabricate and measure lumped-element superconducting resonant circuits that contain either as-deposited or annealed \SiN{}. The loss attributed to the \SiN{} is well-characterized by the standard tunneling model of two-level systems (TLS) supplemented with a self-heating model, which is particularly relevant for the as-deposited \SiN{} devices at large drive strengths. The TLS model includes damping caused both by resonant energy exchange with TLS and by the relaxation of non-resonant TLS \cite{anderson_anomalous_1972,phillips_two-level_1987}. Annealing remedies the power-induced loss at large drive strengths, reduces the relaxation-type damping by more than a hundred fold, and reduces the resonant-type damping by a factor of three. Fourier-transform infrared spectroscopy (FT-IR) reveals that annealing reduces the atomic hydrogen percentage by at least an order of magnitude, implying that atomic hydrogen impurities form TLS that induce microwave-frequency dielectric loss.


\section{Theory of Dielectric Loss}
Amorphous materials can host defect states wherein an electron, atom, or group of atoms within the solid can tunnel between two nearly degenerate configurations, forming a two-level system \cite{muller_towards_2019}. The standard tunneling model describes how the electric and elastic dipole moments of these TLS couple to external electric or strain fields \cite{anderson_anomalous_1972, phillips_two-level_1987, maccabe_nano-acoustic_2020}. In particular, the coupling between the electric dipole moments of a bath of TLS and the electric field of an electrical resonator adds dielectric loss to the circuit.

The TLS model predicts damping via two different coupling mechanisms: transverse and longitudinal. 
Transverse coupling, known as resonant damping, samples TLS whose detuning from the electrical resonance frequency $\omega_0$ is small compared to the TLS decoherence rate. These TLS directly exchange energy with the resonator, and the loss inherited from this interaction can be reduced by increasing the temperature $T$ or drive strength (parameterized by the intracavity photon number \nint{}).
The resonant contribution to the loss is
 \begin{equation}
     Q^{-1}_\text{res}(\overline{n}, T) = \frac{ F_\text{SiN} \text{tan} \delta_\text{0,res}}{\sqrt{1 + \overline{n}/n_c}} \tanh \left(\frac{\hbar \omega_0}{2 k_B T}\right),
     \label{Qresonant}
 \end{equation}
where $n_c$ is a critical photon number describing TLS saturation, and $\tan \delta_\text{0,res}$ is the loss tangent of the dielectric from the resonant interaction at zero temperature. \pSiN{} is the fraction of electric field energy of the electrical resonator mode that is present in the \SiN{} volume.

Longitudinal coupling, known as relaxation damping, does not involve direct energy exchange with the electrical resonator. Instead, the oscillator amplitude shifts the TLS' energy levels such that TLS with energies on the order of the thermal energy are brought out of equilibrium. The TLS then dissipate energy to phonons as they relax back to equilibrium. The loss inherited by the electrical resonator depends on the characteristic TLS relaxation rate, which increases monotonically as a function of temperature. As a result, this loss mechanism is independent of drive strength, but predicts an increase in loss with increased temperature.
When the TLS relaxation rate is small compared to the electrical resonator frequency (generally true for $T \lesssim 1$ K at microwave frequencies \cite{phillips_two-level_1987}), we may approximate the relaxation damping contribution to the loss as a power law
\begin{equation}
    Q^{-1}_\text{rel}(T) = F_\text{SiN} \tan \delta_\text{0,rel} {\left(\frac{T}{T_0}\right)}^d,
    \label{Qrelaxation}
\end{equation}
where $\tan \delta_\text{0,rel}$ is the loss tangent of the dielectric from relaxation damping relative to a reference temperature $T_0$, and $d$ represents the dimensionality of the phonon bath into which the TLS decay (see Appendix \ref{appendix:relaxationTLSLoss} for the full expression) \cite{wollack_loss_2021}. 

Including the TLS damping mechanisms, we describe the total loss with four contributions
\begin{equation}
    Q^{-1}_\text{i}(\overline{n}, T) = Q^{-1}_{\text{bg}} + Q^{-1}_\text{res}(\overline{n}, T) + Q^{-1}_\text{rel}(T) + Q^{-1}_\text{qp}(T),
    \label{Qfull}
\end{equation}
where $Q^{-1}_\text{bg}$ is a constant value of background loss and $Q^{-1}_\text{qp}(T)$ is a parameter-free calculation of the loss from thermal quasiparticles in the superconductor \cite{mattis_theory_1958}. 

We can extend this model by accounting for any potential self-heating of the device. In the presence of a microwave drive, energy is dissipated by the internal loss of the electrical circuit, which can elevate the effective temperature $T_\text{eff}$ of the TLS bath by an amount $T_\text{heat}$ above the base plate temperature of the dilution refrigerator $T_\text{bp}$,
\begin{equation}
    T_\text{eff}(\overline{n}) = T_\text{bp} + T_\text{heat}(\overline{n}).
    \label{self-heating-math}
\end{equation}
In Section IV, we find that $T_\text{heat}$ likely describes the temperature of the TLS bath in the \SiN{}. This effect can manifest as excess loss as a function of drive power in electrical resonators with large self-heating. 

\section{Experimental Design}

\begin{figure}[t]
    \centering
    \includegraphics[width=\linewidth]{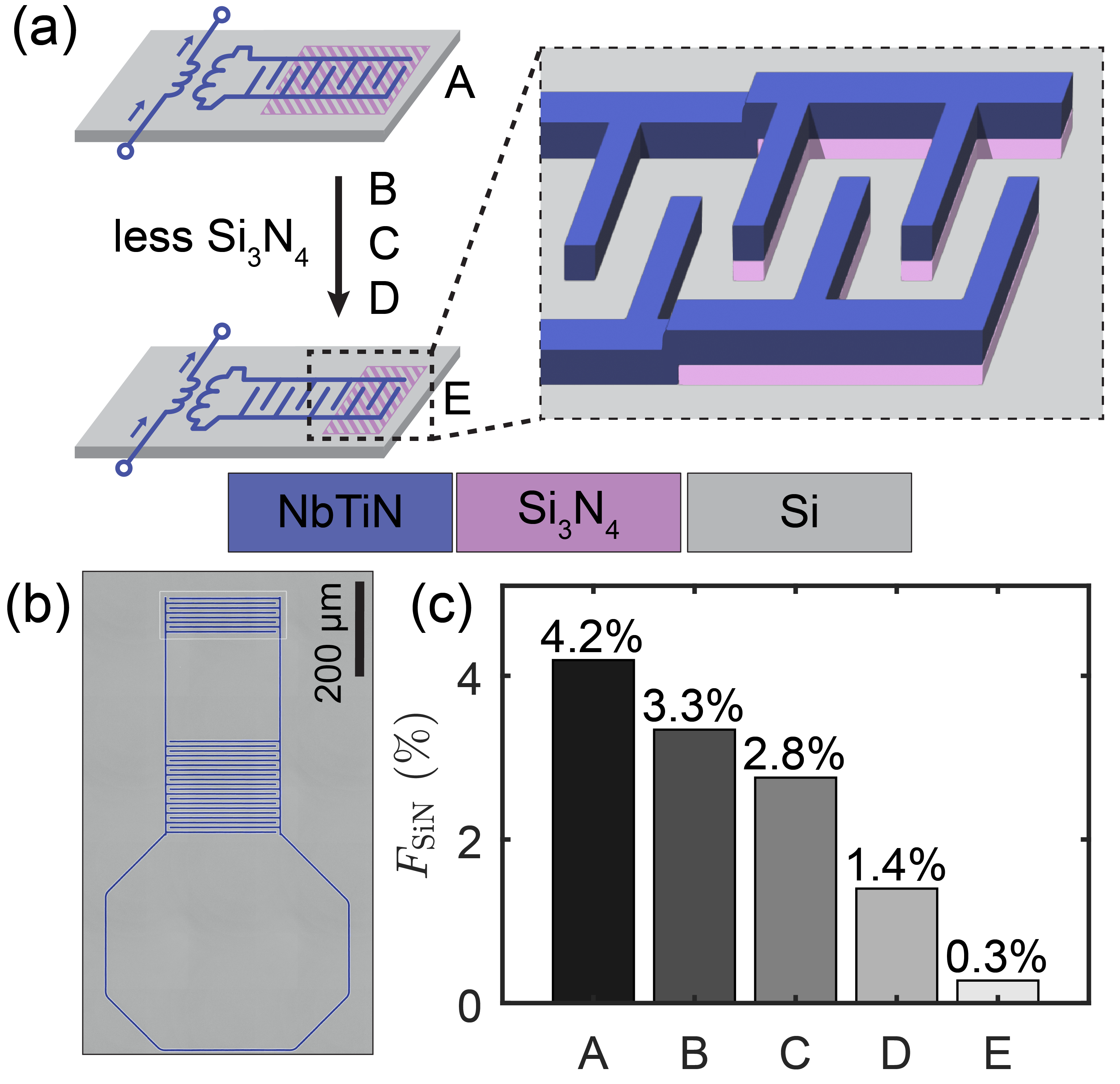}
    \caption{
    \textbf{Circuit schematics and \SiN{} participation.}
    (a) \label{schematics} Schematic representations of the different electrical resonator geometries measured in this work. A NbTiN (blue) readout line couples to a superconducting microwave circuit that partially lies above a \SiN{} film (pink shaded area), all on a silicon chip (gray). The \SiN{} thin film is depicted as a shaded rectangle for clarity in the schematic, but in actual devices, the \SiN{} is only underneath the superconductor, as clarified by the expanded view. (b) \label{image} Example optical microscope image of resonator geometry D. The upper set of capacitor fingers has \SiN{} underneath, while the lower set does not. (c) \label{psin} Electric field energy participation in the \SiN{} for different resonator designs expressed as a percentage. 
    }
\end{figure}

\begin{figure*}[t]
    \centering
    \includegraphics[width=\linewidth]{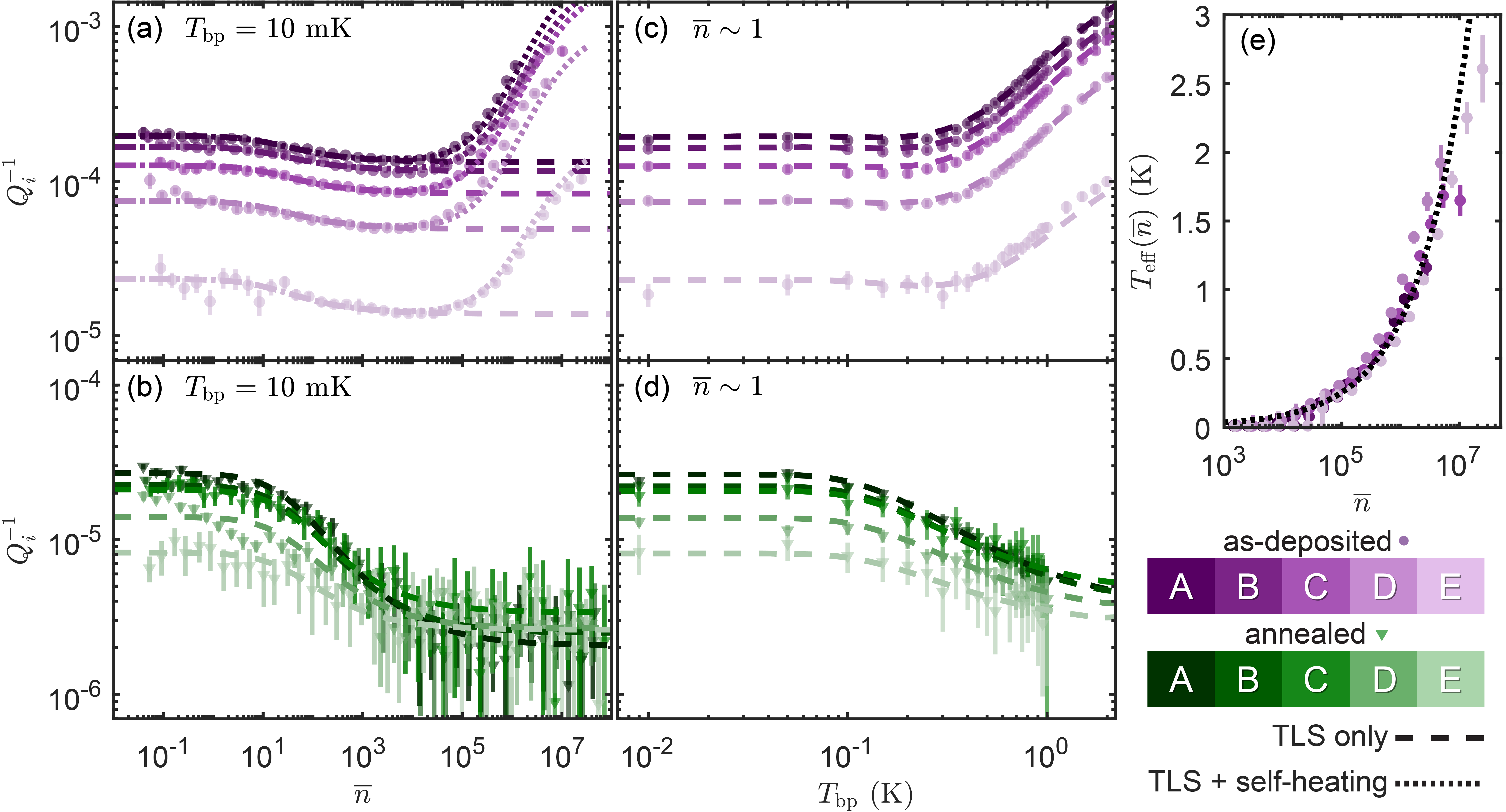}
    \caption{
    \textbf{Power- and temperature-dependent loss of \SiN{}.} \label{Q}
    Loss data for as-deposited \SiN{} (purple circles) and annealed \SiN{} (green triangles) with $\pm1\sigma$ error bars. Different shades represent the different designs A-E. (a-b)\label{Q_n} Loss as a function of drive power at $T_\text{bp} = 10$ mK. The TLS-only model (dashed lines) predicts the decrease of loss as increasing power saturates the TLS absorbers. The combined TLS and self-heating model (dotted lines) accounts for the increase in loss for the electrical resonators with as-deposited \SiN{} at higher powers. (c-d) \label{Q_temp} Loss as a function of base plate temperature at $\overline{n} \sim 1$. For both films, the TLS model captures the behavior without the need to invoke self-heating, due to the low drive power. The increase in loss from relaxation damping observed at higher temperatures with the as-deposited film is absent in the annealed film. (e) \label{selfHeatingFigure} Inferred temperature of the as-deposited \SiN{} as a function of drive power is calculated by correlating the power-dependent and temperature-dependent loss data. The fit (black dotted line) yields the prediction of loss from self-heating in (a).
    }
\end{figure*}

The resonator geometries we use to study the dielectric properties of \SiN{} are represented schematically in Fig. \ref{schematics}(a). The lumped-element circuits are comprised of a loop inductor and interdigitated capacitor. To investigate the dielectric properties of \SiN{}, we design five resonators labeled A to E, with resonance frequencies of about 6 GHz (see Table \ref{appendix:deviceParams}). The key difference in their design is the fraction of capacitor fingers that lie above \SiN{}, with resonator A having the most and resonator E having the least. Fig. \ref{psin}(c) shows the electric field participation ratio \pSiN{} for different devices, which is calculated from finite element simulations in Ansys HFSS \cite{mcrae_materials_2020, wisbey_effect_2010}.

To fabricate the planar circuits, we start by depositing a thin film of \SiN{} on a high-resistivity silicon wafer with a Tystar TYTAN low-pressure chemical vapor deposition (LPCVD) furnace. Using the precursor gasses ammonia (NH$_3$) and dichlorosilane (SiCl$_2$H$_2$) in a 3:1 ratio, we deposit a 100 nm film of \SiN{} at 770$^\circ$C. Post-deposition, we measure a film stress of around 1 GPa and a refractive index at 623 nm of 2.1. These measurements are consistent with a 3:4 stoichiometric ratio of Si and N \cite{paik_reducing_2010}. 
We leave some films as-deposited while for others we perform a four hour anneal of the \SiN{} at 1100$^\circ$C in a pure O$_2$ atmosphere. The atmosphere of O$_2$ is recommended by the furnace manufacturer to drive out the hydrogen impurities better than N$_2$ or vacuum.

We lithographically pattern and etch the \SiN{} film to control the \SiN{} participation of each resonator.
We then sputter a 140 nm film of NbTiN and lithographically pattern the lumped element circuit. The NbTiN etch also removes the remaining \SiN{} that is not masked by the superconductor (expanded view in Fig. \ref{schematics}(a)). 
The relatively thick superconducting layer ensures step coverage of the \SiN{} and minimizes the otherwise strong kinetic inductance nonlinearity associated with thin film NbTiN \cite{giachero_characterization_2023}. Our NbTiN has a superconducting critical temperature $T_c$ of 15 K, so the loss contribution from thermal quasiparticles is sub-dominant to the dielectric loss from TLS at the temperatures studied in this work.

To study the power- and temperature-dependent loss, we excite the resonant circuits with a variable-strength microwave drive tone detuned by 1.5 MHz, while measuring the electrical resonator response using a network analyzer emitting a probe tone with at most 1\% of the power of the drive. Each design is inductively coupled to a common on-chip transmission line, and the device is mounted to the base plate of a dilution refrigerator and cooled to a base temperature of $T_\text{bp} = $ 10 mK. The drive and probe tones access the device through an array of attenuators and cables in the fridge whose microwave transmission response is measured prior to cooling down so that we may calibrate the intracavity photon number \nint{}. To measure the temperature-dependent loss, we hold the drive power fixed and vary the base plate temperature. We fit the frequency-dependent transmission response to extract the resonator's internal loss \inverseQint{}, external coupling \inverseQext{}, and center frequency $\omega_0$ \cite{khalil_analysis_2012, probst_efficient_2015} (Appendix \ref{appendix:circleFit}).

\section{Results and Discussion}

Measurements of the power- and temperature-dependent internal loss of the electrical resonators are shown in Fig. \ref{Q}, along with fits to the model of Eq. \ref{Qfull} with (dotted lines) and without (dashed lines) self-heating. The resonators' internal loss as a function of both power and temperature is directly proportional to electric field participation \pSiN{} (see Fig. \ref{Q_vs_FSiN}). This trend indicates that the power- and temperature-dependent loss originates from the \SiN{} dielectric. Similar circuits with no \SiN{} or as-deposited \SiN{} only under the inductive part of the circuit 
do not demonstrate this behavior (Appendix \ref{appendix:refresAndHalfL}). Qualitatively, we also see that the electrical resonators fabricated on annealed \SiN{} have significantly less loss and exhibit starkly different behavior than that of the resonators with as-deposited \SiN{}. In the following analysis, we begin by applying the TLS-only model, before considering self-heating as a way to understand the full data set.

In Fig. \ref{Q_n}(a-b), we plot the loss as a function of power at base temperature $T_\text{bp}=10$ mK. For the resonators with as-deposited \SiN{}, we observe an initial saturation of the loss with increasing power for \nint{} $\leq 10^4$, followed by a sharp increase at higher powers. This is the same behavior that limited the performance of electromechanical resonators with as-deposited \SiN{} (Appendix \ref{appendix:refresAndHalfL}). Notably, this increase is absent for the resonators with annealed \SiN{}. We can extract the resonant damping contribution $Q^{-1}_\text{res}$ from this low-power data since it is the only power dependent term in Eq. \ref{Qfull}, in the absence of self-heating. From this fit to the TLS-only model, we extract the resonant damping loss tangents $\tan \delta_\text{0,res} = (1.4\pm0.1) \times 10^{-3}$ for the as-deposited film and $\tan \delta_\text{0,res} = (4.8\pm0.4) \times 10^{-4}$ for the annealed film. Annealing yields a factor of three improvement in the low-power, low-temperature dielectric loss. 

The temperature-dependent loss at fixed $\overline{n} = 1$ in Fig. \ref{Q_temp}(c-d) provides information about the relaxation-type loss. For the as-deposited film, we observe an initial saturation of the resonant-type loss with temperature ($T_\textrm{bp} \leq$ 0.3 K), after which the relaxation-type contribution becomes the dominant term ($T_\textrm{bp} \geq$ 0.5 K). Similar to the power-dependent data, the temperature-dependent increase in loss from relaxation damping is absent for the resonators with annealed \SiN{}. No self-heating contribution is required to capture the temperature-dependent data due to the small drive power. Empirically, we find the power law relationship $Q_i^{-1} \propto T^2$ in the as-deposited \SiN{} resonators between 0.3 K and 1 K. Accordingly, we fix $d=2$ in the full expression of $Q^{-1}_\text{rel}(T)$ (Eq. \ref{Qrelaxation_full}) for the remaining analysis. The contrast between the increasing loss with temperature for as-deposited \SiN{} and the lack thereof for annealed \SiN{} is reflected in the extracted loss tangents. We find $\tan \delta_\text{0,rel} = (3.4\pm0.1) \times 10^{-3}$ for the as-deposited film and $\tan \delta_\text{0,rel} = (8\pm8) \times 10^{-6}$ for the annealed film with reference temperature $T_0 = 500$ mK, representing a dramatic decrease of this loss mechanism.

Interestingly, annealing realizes only a modest improvement to the resonant-type loss, but vastly reduces the relaxation-type loss. As noted in Section II and further explored in Appendix \ref{appendix:relaxationTLSLoss}, the resonant interaction couples the electrical resonator to TLS with energy close to $\omega_0$, while the relaxation interaction mainly couples to off-resonant TLS. The relaxation contribution starts to dominate over the resonant contribution above $T_\text{bp} \sim 500$ mK, suggesting that the impurities removed by annealing were host to TLS with energies predominantly above $500~\text{mK} \times 2 k_B/h \simeq 20$ GHz. This observation also suggests that the annealing changes the frequency-dependent structure of the TLS density of states.

Although the drastic decrease in the relaxation-type loss may seem inconsequential for experiments at $T \ll T_0$, this loss mechanism can become relevant in situations where a high-power drive tone induces local heating. We propose that the anomalous increase in loss at high drive strengths observed in the resonators with as-deposited \SiN{} seen in Fig. \ref{Q_n}(a) could be a symptom of self-heating, which activates the temperature-dependence of the relaxation-type loss. The self-heating mechanism is suggested by the striking qualitative similarity between the power- and temperature-dependent loss of as-deposited \SiN{}. A similar qualitative agreement is observed between the power- and temperature-dependent electrical resonator frequency shifts (see Appendix \ref{appendix:freqShiftTheory}). We can parameterize the power-dependent loss within the TLS model by assigning an effective temperature $T_\text{eff}$ to the electrical resonator at each drive power $\overline{n}$. We use the fits to the TLS-only model of the temperature-dependent loss to determine the temperature that is consistent with the measured dissipation $Q^{-1}(\overline{n}, T_\text{eff})$.

This mapping for the electrical resonators with as-deposited \SiN{} is shown in Fig. \ref{selfHeatingFigure}(e). Though the mapping for each device A-E is made separately, the inferred temperature curves obey the same functional dependence with $\overline{n}$ independent of \pSiN{}. Resonators with more \SiN{} dissipate more power in proportion to the relative amount of \SiN{}, but the dielectric also has more thermalization pathways due to the larger area. As a result, the inferred temperature is about the same for a given \nint{}, despite the differences in geometry. We can perform a self-consistency check of the TLS and self-heating model by plotting the power-dependent loss predicted by Eq. \ref{Qfull}, now including a power-dependent temperature. We parameterize the effective temperature by jointly fitting the data from all the resonators in Fig. \ref{selfHeatingFigure}(e) to a simple power-law: $T_\text{eff}(\overline{n}) = A \overline{n}^\beta + T_\text{bp}$. Empirically, we find $\beta \approx 0.5$, and subsequently fix this exponent in the fitting. The consequence of this analysis is shown in Fig. \ref{Q}(a) in the dotted lines, demonstrating that self-heating could explain the increase in loss at high powers in the as-deposited films.

This analysis suggests significant heating of the relevant degrees of freedom, up to more than 2 K with \nint{} $ = 10^7$. In principle, $T_\text{eff}$(\nint{}) could reflect the temperature of either the whole Si chip, the \SiN{} film, or the bath of TLS. However, we observe no power-induced loss when driving one circuit and measuring the loss of another circuit on the same chip, indicating the Si chip remains cold. Likewise, calculations and observations of the thermal boundary resistance at the \SiN{}-Si interface \cite{swartz_thermal_1989, lee_heat_1997} are inconsistent with the temperature difference we appear to measure. This leaves the possibility that the TLS bath itself is at an elevated temperature, which can occur if the TLS-TLS interactions are stronger than the TLS-phonon interaction. In the as-deposited films, the TLS-TLS interaction may be significantly enhanced relative to the annealed films due to the large density of impurities \cite{black_spectral_1977}, as reported in the next section.


\begin{figure}[t]
    \centering
    \includegraphics[width=\linewidth]{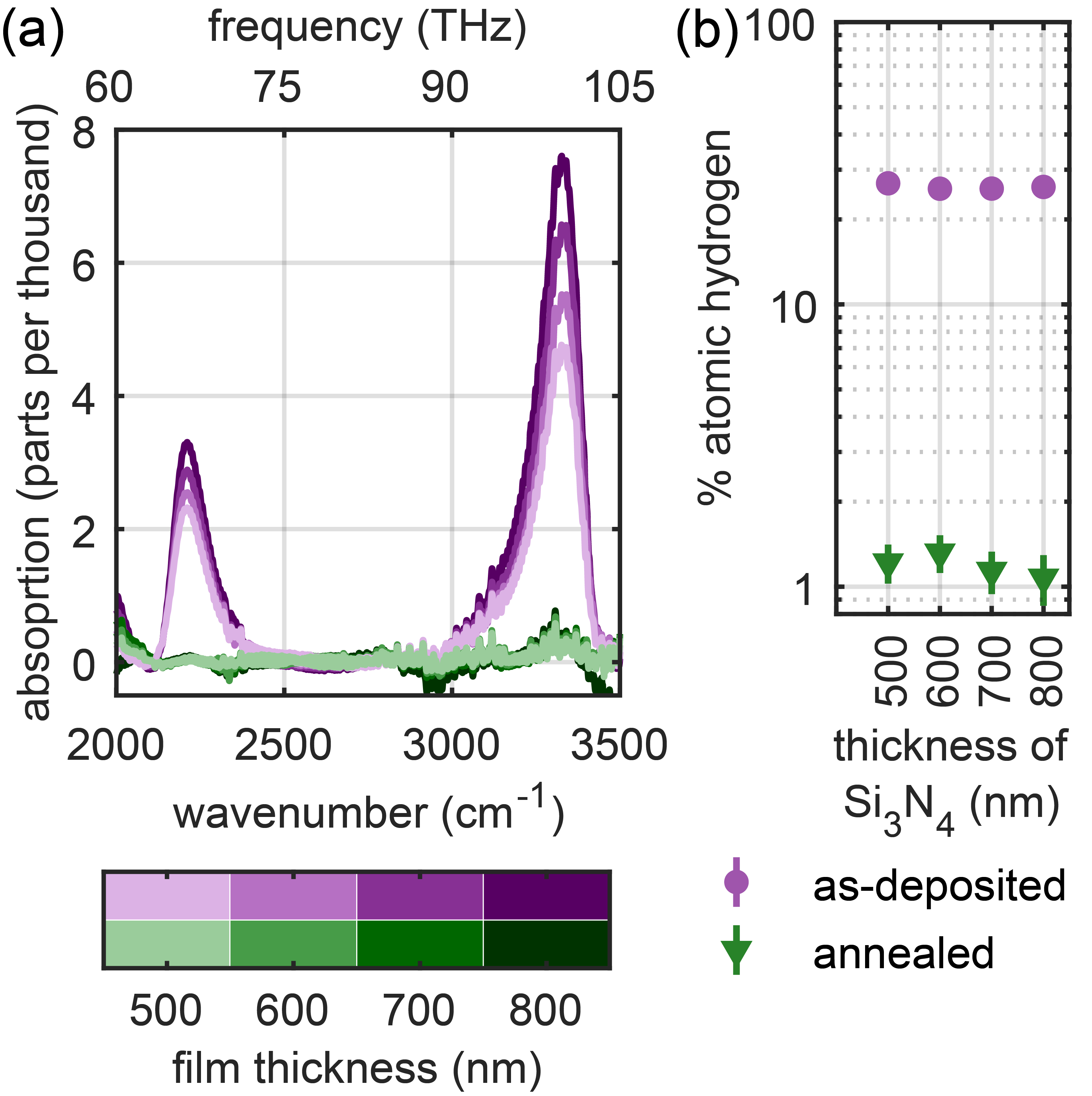}
    \caption{
    \textbf{Infrared spectroscopy reveals hydrogen impurities.} (a)\label{spectrum} The N-H and Si-H bond stretching vibrational modes are in the IR spectrum between 60 and 100 THz at wavenumbers of 3330 cm$^{-1}$ and 2210 cm$^{-1}$, respectively. We measure \SiN{} films of different thicknesses with Fourier-transform infrared spectroscopy (FT-IR) before and after annealing, noticing a significant reduction in the absorbance at these wavenumbers. (b)\label{hpercent} The atomic hydrogen percentage in the film is calculated from the spectrum measurements. The annealed film has more than an order of magnitude less hydrogen content than the as-deposited film. Error bars represent $\pm1\sigma$. 
    }
    \label{FTIR}
\end{figure}

\section{Microscopic Origin of TLS}

To better understand the impact of annealing on the dielectric loss of \SiN{}, we use Fourier-transform infrared spectroscopy (FT-IR) to attempt to identify the microscopic source of the TLS. For our \SiN{} film, we expect to primarily find molecular vibrational modes of the Si-N bond and Si-H or N-H impurities \cite{paik_reducing_2010, herth_optimization_2010}. As-deposited and annealed \SiN{} films of various thicknesses are prepared (details in Appendix \ref{appendix:FTIR_details}), and results of the FT-IR measurements are shown in Fig. \ref{spectrum}(a).

The spectra reveal that the as-deposited \SiN{} is host to hydrogen impurities, whose atomic concentration is substantially reduced after annealing. In addition to the Si-N stretching mode at 890 cm$^{-1}$ (not shown), we observe peaks in the absorption spectrum at 3330 cm$^{-1}$ and 2210 cm$^{-1}$ for the as-deposited films corresponding to the stretching modes of the N-H and Si-H bond, respectively \cite{paik_reducing_2010}. The height of these peaks decreases with decreasing film thickness.  Notably, the peaks of the spectrum for the annealed films are nearly indistinguishable from the background. By comparing the atomic hydrogen percentage in the films \cite{lanford_hydrogen_1978}, Fig. \ref{hpercent}(b) shows that annealing reduces the hydrogen content of the \SiN{} by at least an order of magnitude. Additionally, the hydrogen content for both films is independent of thickness, implying the impurities are uniformly distributed throughout the bulk.

The near-IR modes identified by the FT-IR are likely too high in frequency to be the source of loss in the \SiN{}, but they do serve as markers of hydrogen impurities that can host other TLS. Hydrogen impurities in alumina (Al$_2$O$_3$) are strong candidates for GHz-frequency TLS \cite{holder_bulk_2013, muller_towards_2019}, where an interstitial hydrogen atom tunnels between adjacent oxygen atoms \cite{gordon_hydrogen_2014}. Analogous defects in the 20-50 GHz range have been found in niobium films, where interstitial hydrogen atoms tunnel between adjacent nitrogen impurities \cite{morr_isotope_1989, steinbinder_quantum_1991, cannelli_tunneling_1998}. Similar defects could exist in our \SiN{} films. While the data presented in this work do not conclusively identify a specific physical TLS realization, it does suggest that the source of the TLS is intimately related to the presence of hydrogen impurities, which we have shown are removed by thermal annealing.

\section{Conclusion}

We have investigated the power- and temperature-dependent dielectric properties of \SiN{} with lumped-element superconducting resonators. We observe not only the familiar saturable loss associated with the resonant interaction between TLS and electrical resonator, but also a significant temperature-dependent loss that is attributed to relaxation-type damping. The as-deposited \SiN{} films also exhibit increasing power-induced loss at large powers, which we understand via a combined TLS and self-heating model. We show that high-temperature annealing reduces the relaxation damping component significantly, while also modestly improving the resonant damping contribution. Moreover, the increasing power-induced loss at large powers observed in the as-deposited films is absent in the annealed films. There have been similar anomalous power-dependence observations in other materials, such as amorphous silicon \cite{shu_nonlinearity_2021}, crystalline silicon \cite{checchin_measurement_2022}, and sapphire \cite{read_precision_2023}, which could also originate from TLS relaxation damping activated by local self heating. In \SiN{}, we attribute the source of dissipation to hydrogen impurities in the dielectric, which are reduced by an order of magnitude in the annealing process.

Our combined TLS and self-heating model also highlights the importance of proper thermalization. Systems with suspended \SiN{} are particularly vulnerable to self-heating, as the pathways for thermalization are already restricted \cite{zhang_radiative_2020, ftouni_thermal_2015}. In fact, the power-induced loss observed in electro-optomechanical transducers was larger than that measured in these planar circuits despite having smaller \pSiN, suggesting considerable heating of the dielectric (see Appendix \ref{appendix:converter}) \cite{brubaker_optomechanical_2022, delaney_superconducting-qubit_2022}. In the absence of power-induced internal loss from relaxation damping, the electromechanical interaction, which is mediated by a strong drive tone, can be maximized without negatively impacting the resonant circuit. 

Another limiting factor of the electro-optomechanical transducer's performance was microwave noise induced by the strong microwave drive tone. From the fluctuation-dissipation theorem, a portion of this noise must have resulted from the loss and self-heating identified in this work. Further investigation could interrogate the degree to which annealing also improves the power-induced microwave noise.

The data collected for this study and the code used to generate the figures for this paper are available on Zenodo \cite{mittal_zenodo_2023}.

\section{Acknowledgements}
We thank James Beall and Michael Vissers for helping with device fabrication, and Jordan Wheeler for helping with $T_c$ measurements of our NbTiN films. Benjamin Brubaker, Jonathan Kindem, Peter Burns, and Andrew Higginbotham launched this study with early investigations of the loss and noise in \SiN{} electromechanical devices. Additionally, we thank Shuo Sun and Jacob Davidson for helpful discussions. This work was supported by funding from ARO Grant W911NF2310376, NSF Grant No. PHY-2317149, NSF QLCI Award OMA - 2016244, the Office of the Secretary of Defense via the Vannevar Bush Faculty Fellowship, award No. N00014-20-1-2833, and the Baur-SPIE Endowed Chair at JILA.

\appendix
\renewcommand{\thefigure}{\thesection\arabic{figure}}
\renewcommand{\theequation}{\thesection\arabic{equation}}
\renewcommand{\thetable}{\thesection\arabic{table}}
\setcounter{figure}{0}
\setcounter{equation}{0}
\setcounter{table}{0}

\section{Determining Resonator Parameters}
\label{appendix:specs}
\label{appendix:circleFit}
Electrical resonators A-E are inductively coupled to a single transmission line that traverses the chip, and we measure the transmission response in order to extract the resonators' parameters. Each resonator is designed to have a unique frequency both from varying \SiN{} participation and from small variation in the designed inductance. We first preprocess the frequency-dependent data by normalizing the off-resonance values to unity transmission. We then fit the response to 
\begin{equation}\label{eq:S11}
    S_{21}(\omega) = 1 - \frac{Q |Q_e^{-1}| e^{i\phi}}{1 + 2iQ \frac{\omega - \omega_0}{\omega_0}}
\end{equation}
to extract the quality factor of the resonator $Q$, the external quality factor $Q_e$, and center frequency $\omega_0$. 
We use a complex external quality factor $Q_e^{-1} = |Q_e^{-1}| e^{i\phi}$ to account for unintended reflections that lead to asymmetry in the frequency response \cite{khalil_analysis_2012, probst_efficient_2015}. We define the internal quality factor as $Q_i^{-1} = Q^{-1} - \Re\{Q_e^{-1}\}$. 


Using the fitted resonator parameters and known attenuation of the fridge input line, we calculate the intracavity photon number \nint{} using
\begin{equation}\label{eq:nint}
    \overline{n} = \frac{P_\text{inc}}{ \hbar \omega_0^2} \frac{ \frac{1}{2}|Q_e^{-1}|}{(\frac{1}{2}Q^{-1})^2 + (\Delta_p/\omega_0)^2},
\end{equation}
where $P_\text{inc}$ is the incident power on the resonator and $\Delta_p = \omega_p - \omega_0$ is the detuning of the drive tone from the resonator center frequency ($\Delta_p/2\pi = 1.5$ MHz in this work). Values for the fitted parameters at $T_\text{bp} = 10$ mK and \nint{} $\sim 1$ are shown in Table \ref{appendix:deviceParams} for both the as-deposited and annealed resonators.

\begin{table}[t]
\centering
\begin{tabular}{|c|c|c|c|c|} \hline
& Device & $\omega_0/2\pi$ (GHz) & $Q_i^{-1} \times 10^{5}$ & $|Q_e^{-1}| \times 10^{5}$ \\ 
\hline 
\multirow{5}{*}{as-deposited} & A & 5.968 & 18.4 & 9.3 \\ 
& B & 6.133 & 16.2 & 10.3 \\ 
& C & 6.289 & 12.5 & 22.4 \\ 
& D & 6.384 & 7.5 & 9.8 \\ 
& E & 6.480 & 1.9 & 15.3 \\ \hline
\multirow{5}{*}{annealed} & A & 5.959 & 2.4 & 9.8 \\ 
& B & 6.103 & 2.2 & 12.4 \\ 
& C & 6.271 & 1.9 & 23.1 \\ 
& D & 6.362 & 1.3 & 8.7 \\ 
& E & 6.443 & 0.8 & 16.6 \\ 
\hline 
\end{tabular}
\caption{\textbf{Device parameters.} \label{appendix:deviceParams} The center frequency, internal quality factor, and external quality factor for the studied devices measured at $T_\text{bp} = 10$ mK and $\overline{n} \sim 1$. For all resonators, the external quality factor is roughly within an order of magnitude of the internal loss, which allows for accurate determination of the relative coupling rates from the numerical fits.}
\end{table}

\section{Scaling of Loss with \SiN{} in Related Electrical Resonators}
\setcounter{figure}{0}
\label{appendix:refresAndHalfL}
\label{appendix:converter}

\begin{figure}[t]
    \centering
    \includegraphics[width=\linewidth]{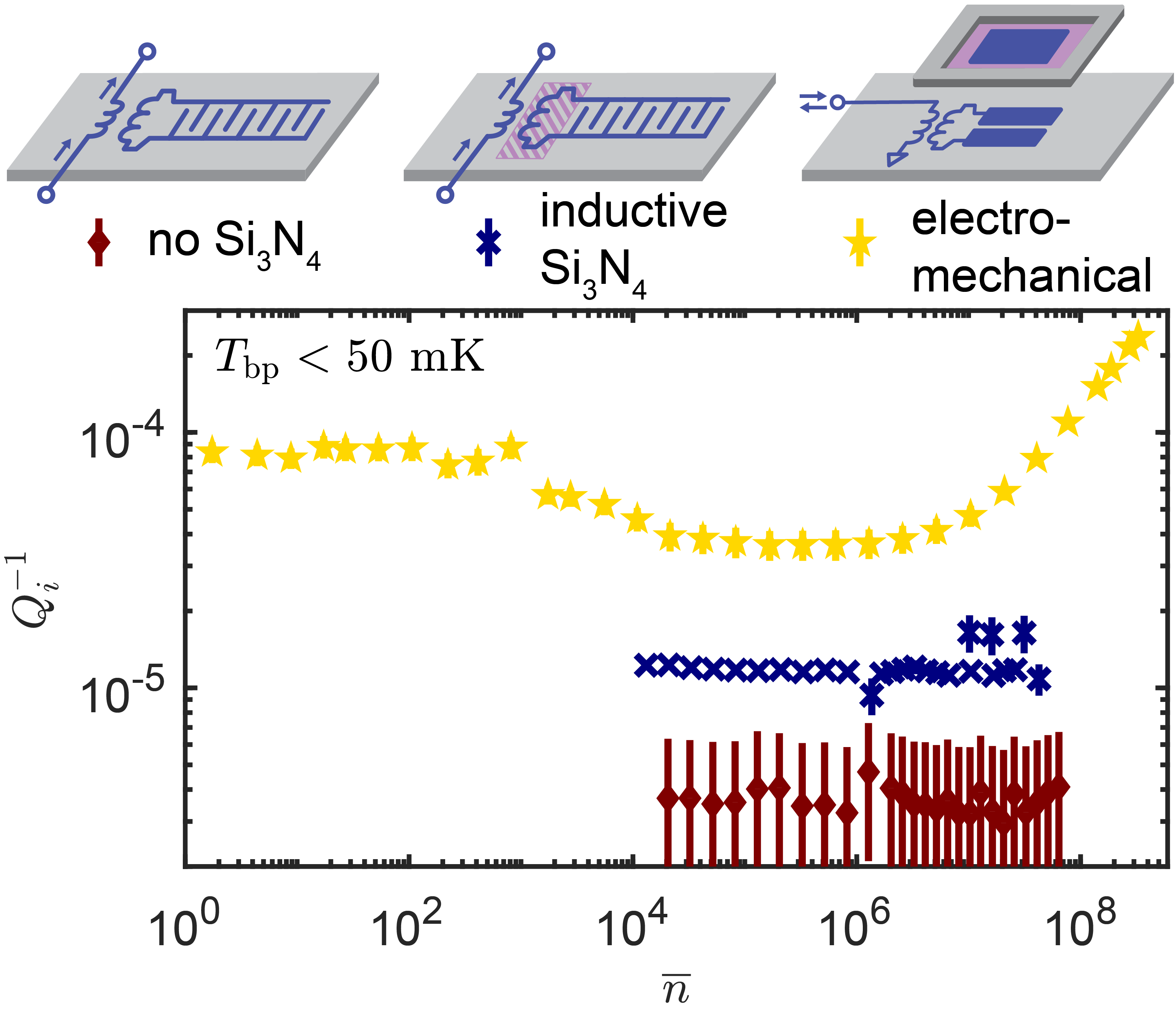}
    \caption{
        \textbf{Power-dependent loss in electrical resonators with different geometries.}\label{otherResonators} Resonators with no \SiN{} (red diamonds) or \SiN{} under the inductor (blue crosses) have loss which saturates to low values at moderate drive strengths $\overline{n} \sim 10^4$. The loss of a device with electromechanically compliant \SiN{} (yellow stars) saturates to a higher value and furthermore rapidly increases with larger drive strengths, likely due to poor thermalization. Error bars indicate $\pm1\sigma$ uncertainty for the numerical fits.
    }
\end{figure}

In order to test whether the power- and temperature-dependent loss could be caused by something other than dielectric loss, we fabricated devices either with no \SiN{} or with \SiN{} only under the inductor rather than the capacitor in the circuit. Schematic depictions and measurements of these devices are shown in Fig. \ref{otherResonators}. \pSiN{} is 0 for the resonator with no \SiN{}, and \pSiN{} is 0.002\% for the resonator with \SiN{} under the inductive part of the circuit -- orders of magnitude lower than resonators A-E. We observe almost no change in the loss with power in these resonators, as expected if the loss in the resonators A-E is from the \SiN{} dielectric. The behavior of these devices shows that other possible sources of loss, such as excess quasiparticles or mobile magnetic vortices, are inconsistent with the observed power-induced loss \cite{mcrae_materials_2020}.

\begin{figure}[t]
    \centering
    \includegraphics[width=\linewidth]{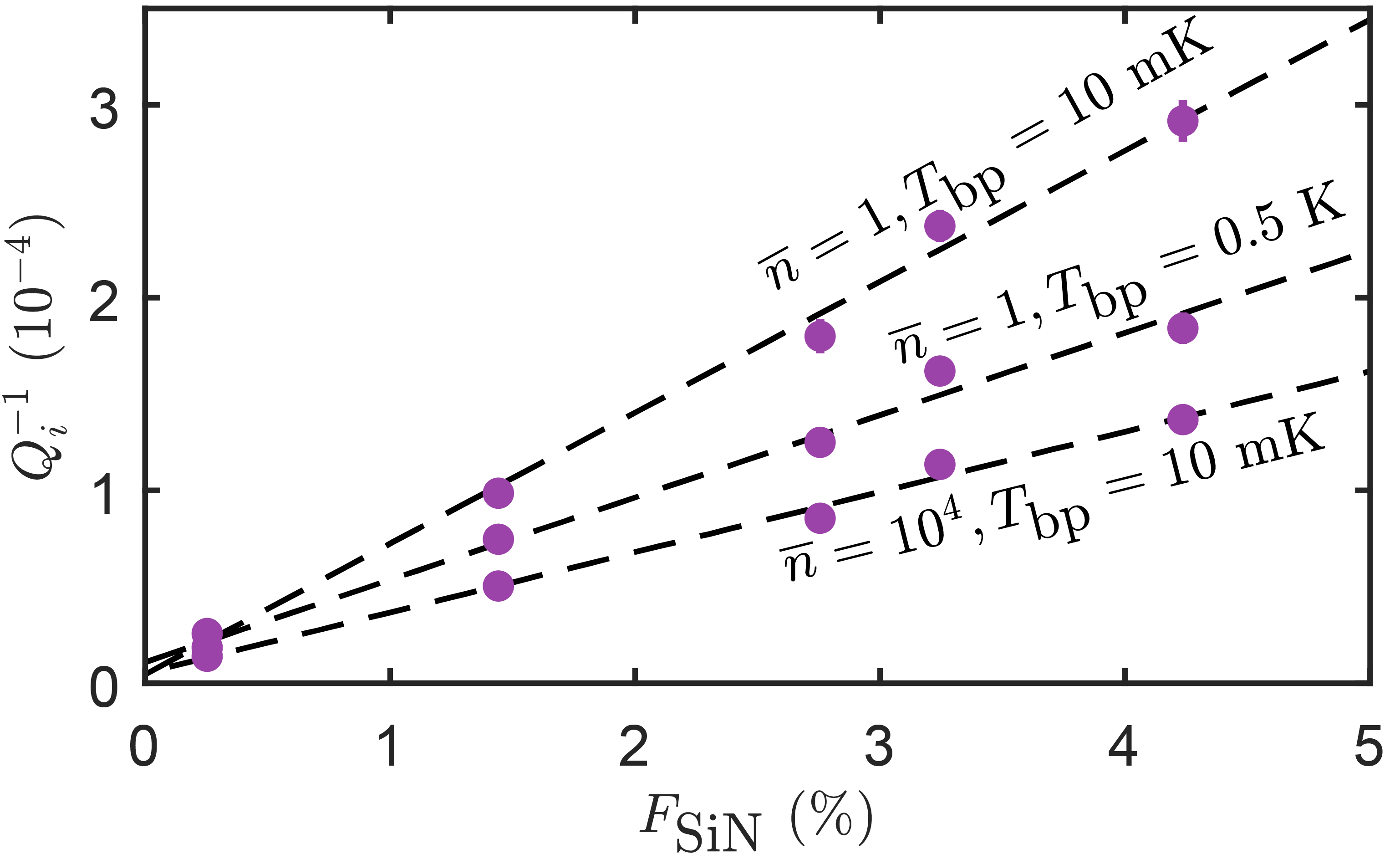}
    \caption{
        \textbf{Loss versus \pSiN{}.}\label{Q_vs_FSiN} The internal loss of the resonators A-E with as-deposited \SiN{} scales linearly with electric field participation \pSiN{}. This relationship holds at different drive strengths and temperatures. Error bars indicate $\pm1\sigma$ uncertainty.
    }
\end{figure}

The power-dependent loss measured in an electromechanical device is also plotted in Fig. \ref{otherResonators} \cite{brubaker_optomechanical_2022}. Due to variability in the geometry from misalignment of the electromechanical capacitor, \pSiN{} can vary, but for reasonable assumptions of the alignment, we simulate \pSiN{} $\sim 0.1\%$. Although this value is less than the planar circuits discussed in the main text, this electromechanical resonator exhibits a very strong increase in the power-induced loss above \nint{} $\sim 10^7$. The power-dependence suggests restricted thermalization pathways for the phonons in the suspended \SiN{} membrane, such that the TLS interacting with a phonon bath at a temperature elevated from that of the base plate of the dilution refrigerator.

As shown in Fig. \ref{Q_vs_FSiN}, we verify that the loss is determined by the dielectric loss of \SiN{} by plotting the internal loss $Q_i^{-1}$ as a function of electric field participation \pSiN{} for as-deposited devices. The loss is linearly proportional to \pSiN{} for a range of drive strengths and temperatures.



\section{TLS Model Predictions of Loss}
\label{appendix:relaxationTLSLoss}
\setcounter{figure}{0}
Following the derivation from Phillips (1987) \cite{phillips_two-level_1987}, the loss contribution from the resonant interaction is
\begin{equation}
    Q^{-1}_\text{res}(\overline{n},T) = \frac{P |\mathbf{\mean{p}_0}|^2 \pi}{3 \epsilon} \frac{\tanh{(\hbar \omega_0/2 k_B T)}}{\sqrt{1 + \overline{n}/n_c}},
    \label{Q_resonant_appendix}
\end{equation}
where $|\mathbf{\mean{p}_0}|$ is the spatially averaged TLS dipole moment, $\epsilon$ is the permittivity of the dielectric, and $P$ is the TLS density of states. In practice, we absorb these numerical quantities into $F_\text{SiN} \tan \delta_\text{0,res}$ in Eq. \ref{Qresonant}.
The contribution to loss from the relaxation interaction is
\begin{multline}
    Q^{-1}_\text{rel}(T) = \frac{P |\mathbf{\mean{p}_0}|^2 \omega_0}{6 \epsilon k_B T} \int_0^{\infty} \dif E \int_{\tau_{\text{min}}(E)}^{\infty} \dif \tau \sech^2\left(\frac{E}{2 k_B T}\right) \\
    \times \sqrt{1 - \frac{\tau_{\text{min}}(E)}{\tau}} \frac{1}{1 + \omega_0^2 \tau^2}.
    \label{Qrelaxation_full}
\end{multline}
The double integral is over the TLS energies $E$ and the TLS relaxation time $\tau$. The TLS decay is dominated by phonon interactions, and the relaxation rate due to one-phonon processes in $d$ dimensions is given by Eq. 7 of Ref. \cite{behunin_dimensional_2016},
\begin{equation}
    \label{taumin}
    \tau^{-1}(E) = \frac{\mean{\gamma}^2}{\mean{v}^{d+2}} \frac{\pi S_{d-1}}{(2 \pi)^d} \frac{E^{d-2} \Delta_0^2}{\hbar^{d+1} \rho_d} \coth \left(\frac{E}{2 k_B T}\right),
\end{equation}
where $\mean{\gamma}$ is the TLS elastic dipole averaged over position and orientation, $\mean{v}$ is the polarization-averaged acoustic velocity, $S_d$ is the $d$-dimensional unit-hypersurface area, $\Delta_0$ is the TLS tunneling strength, and $\rho_d$ is the $d$-dimensional density. The relaxation time is minimized when the TLS energy $E$ matches the tunneling strength $\Delta_0$ \cite{phillips_two-level_1987}, resulting in $\tau_{\text{min}}^{-1}(E) \propto E^d \coth (E/2 k_B T)$. The dimensionality of the phonon bath experienced by the TLS is determined by the temperature and geometry of the dielectric material \cite{hoehne_damping_2010, behunin_dimensional_2016}.

In the high-temperature regime, the TLS lifetime is reduced by higher order phonon processes such that $\omega_0 \tau_{\text{min}} \ll 1$, so we can approximate $\tau_{\text{min}} \rightarrow 0$. In this limit, the loss saturates to a value independent of temperature,
\begin{equation}
    \lim_{\omega_0 \tau_\text{min} \ll 1} Q^{-1}_\text{rel}(T) = \frac{P |\mathbf{\mean{p}_0}|^2 \pi}{6 \epsilon}.
    \label{Qrel_hightemp}
\end{equation}
In the low-temperature regime, when $\omega_0 \tau_{\text{min}} \gg 1$ and the TLS are long lived, Eq. \ref{Qrelaxation_full} reduces to a power law with exponent $d$ described earlier in Eq. \ref{Qrelaxation}. To see this, we can approximate $1/(1 + \omega_0^2 \tau^2) \approx 1/(\omega_0^2 \tau^2)$. Then, the integral over $\tau$ can be evaluated using the identity $\int_a^\infty \sqrt{(1-a/x)}/x^2 = 2/(3a)$. Putting it all together, we have
\begin{multline}
     \lim_{\omega_0 \tau_\text{min} \gg 1} Q^{-1}_\text{rel}(T) = \frac{P |\mathbf{\mean{p}_0}|^2 \omega_0}{9 \epsilon} \frac{\mean{\gamma}^2}{\mean{v}^{d+2}} \frac{S_{d-1}}{(2 \pi)^{d-1}} \frac{1}{\hbar^{d+1} \rho_d} \\ 
     \times (2 k_B T)^d \int_0^{\infty} \xi^d \sech^2(\xi) \coth(\xi) \dif\xi,
     \label{Qrel_lowtemp}
\end{multline}
where we have performed the change of variables $\xi = E/(2 k_B T)$. This expression includes both the TLS' electric and elastic dipole properties because the TLS absorbs electrical energy and subsequently decays to phonons. In Eq. \ref{Qrelaxation}, we absorb all of the numerical quantities into $F_\text{SiN} \tan \delta_\text{0,rel}/T_0^d$ but the fits to the data employ the full expression from Eq. \ref{Qrelaxation_full}.


A key assumption of the standard tunneling model is a TLS density of states $P$ that is constant with respect to energy \cite{phillips_two-level_1987}. Consequently, the model predicts similar scales for the loss inherited from the resonant- and relaxation-type interactions, but we observe very different scales for the two contributions in practice. For example, annealing improves the relaxation-type loss much more than the resonant-type loss. Additionally, the high temperature asymptote of the relaxation contribution (Eq. \ref{Qrel_hightemp}) is significantly larger than the zero-temperature, zero-power limit of the resonant contribution (Eq. \ref{Q_resonant_appendix}) for the devices with as-deposited \SiN{}. The need for independent pre-factors in our fits is further evidence of a non-uniform energy distribution of the TLS, as the resonant and relaxation contributions sample different regions of the TLS density of states.

\begin{figure}[t]
    \centering
    \includegraphics[width = \linewidth]
    {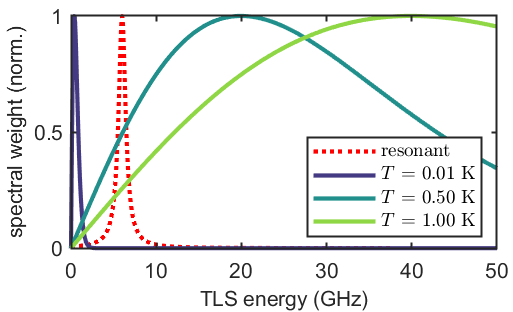}
    \caption{
    \textbf{Sampling of TLS Energy Distribution.} 
    The resonant contribution (dotted line) samples a narrow distribution of energies, centered on the frequency of the microwave circuit ($\sim$ 6 GHz). The relaxation contribution (solid lines) samples different regions of TLS energy at different temperatures. For $T \geq $ 0.5 K, where we start to observe appreciable relaxation damping in the as-deposited films, a wide distribution of TLS with energies $\gtrsim$ 10 GHz are relevant.
    }
    \label{appendix:TLSsampling}
\end{figure}

We plot the regions of the TLS distribution that are being sampled by the two damping terms in Fig. \ref{appendix:TLSsampling}. The resonant contribution is sensitive to TLS with energies close to resonance frequency of the microwave circuit, with the width of the distribution set by the characteristic TLS decoherence rate. To see the TLS energies sampled by the relaxation contribution, we plot the value of the integrand in Eq. \ref{Qrel_lowtemp} as a function of the TLS energy at various temperatures for $d = 2$. This distribution is centered on $E \sim 2 k_\textrm{B} T$, with a characteristic width of $\delta E \sim 1.7 \times 2 k_\textrm{B} T$. Thus, the relaxation damping samples a distribution of TLS energies whose center and width scale proportionally with the temperature. From Fig. \ref{appendix:TLSsampling}, it is clear that for temperatures $T \gtrsim $ 0.5 K, TLS with energy much greater than that of the microwave circuit cause the relaxation damping loss.

\section{TLS Model Predictions of Frequency Shift and Data}
\label{appendix:freqShiftTheory}
\setcounter{figure}{0}


\begin{figure*}[t]
    \centering
    \includegraphics[width=\linewidth]{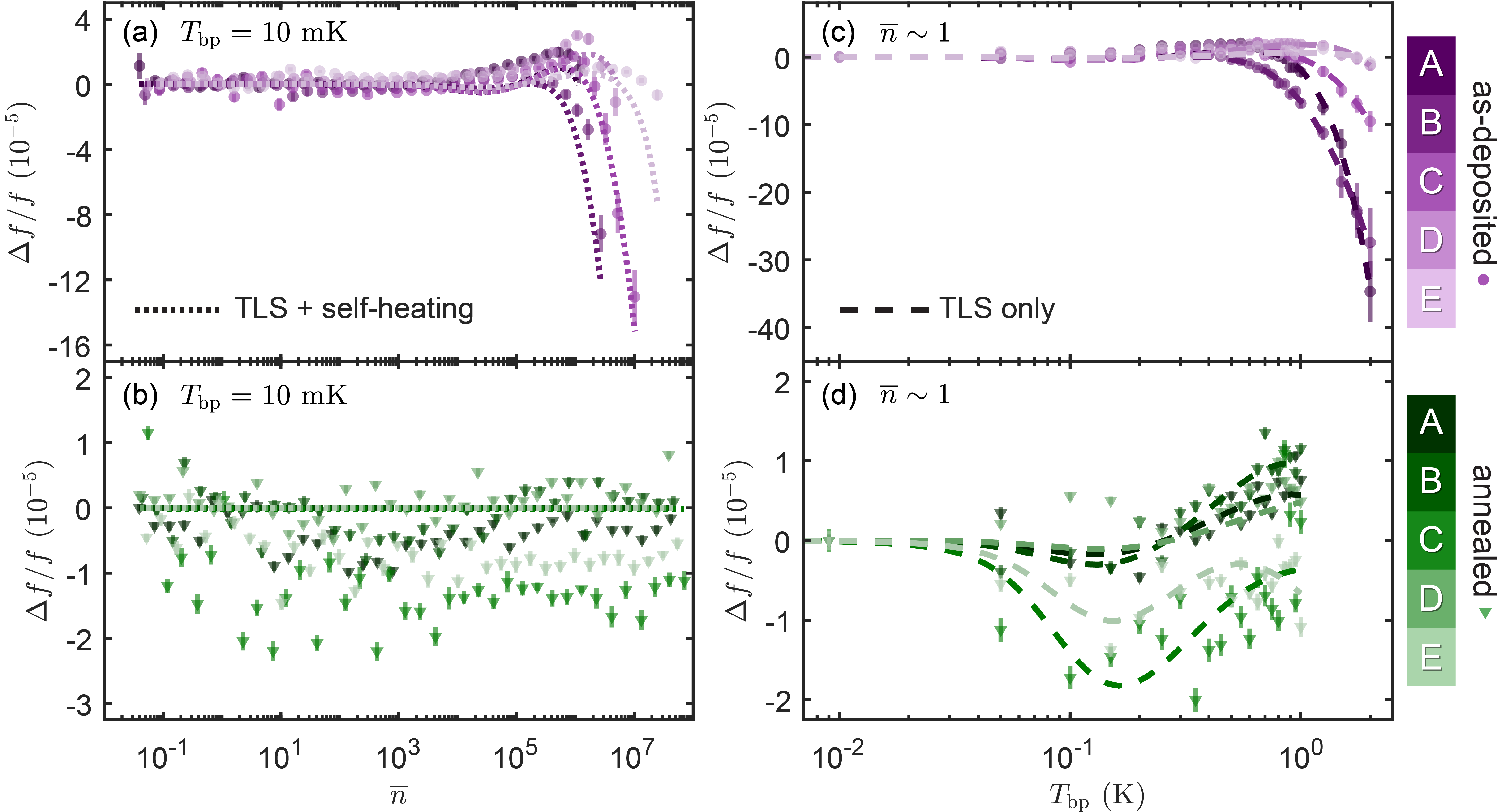}
    \caption{
    \textbf{Power- and temperature-dependent resonator fractional frequency shift}. Frequency shift data for as-deposited \SiN{} (purple circles) and annealed \SiN{} (green triangles) with $\pm1\sigma$ error bars. Different shades represent the different designs A-E. Each resonator's frequency has been normalized with respect to its frequency at base plate temperature $T_\text{bp}=10$ mK and $\overline{n}\sim1$ intracavity photon. (a-b) Power-dependent frequency shift at 10 mK. The TLS model does not predict a frequency shift with drive power, as evidenced by the devices with annealed \SiN{}. Using the self-heating model shown in Fig. \ref{Q} translates power-dependence into temperature-dependence, which captures the qualitative behavior of frequency shift observed for the as-deposited \SiN{}.(c-d) Temperature-dependent frequency shift at $\overline{n}$. The resonators with as-deposited \SiN{} exhibit both resonant and relaxation contributions, while the resonators with annealed \SiN{} only exhibits the resonant contribution, supporting the large reduction in the strength of the relaxation interaction from annealing. 
    }
    \label{freqShift}
\end{figure*}

In addition to a change in the loss rate, we measure the frequency of the resonators as a function of power and temperature. The TLS model predicts frequency shifts from both resonant- and relaxation-type interactions. The contribution from the resonant-type interaction is
\begin{multline}
    \left(\frac{\Delta f}{f}\right)_\text{res} = \frac{P |\mathbf{\mean{p}_0}|^2}{3 \epsilon} \\
    \times \left[ \Re \left\{ \Psi \left(\frac{1}{2} + \frac{\hbar \omega_0}{2 \pi i k_B T}\right)\right\} - \ln \left(\frac{\hbar \omega_0}{2 \pi k_B T}\right) \right],
    \label{freqShiftResonant}
\end{multline}
where $\Psi$ is the digamma function \cite{wollack_loss_2021, maccabe_nano-acoustic_2020}. The first term results in an initial red shift of the resonance frequency as temperature increases, while the second term constitutes a blue shift. 


The relaxation term involves integrating over all the relaxation timescales of the many off-resonant-TLS \cite{phillips_two-level_1987},
\begin{multline}
    \left(\frac{\Delta f}{f}\right)_\text{rel} = \frac{- P |\mathbf{\mean{p}_0}|^2}{12 \epsilon k_B T} \int_0^{\infty} \dif E \int_{\tau_{\text{min}}(E)}^{\infty} \dif \tau \sech^2 \left(\frac{E}{2 k_B T}\right) \\
    \times \frac{1}{\tau} \sqrt{1 - \frac{\tau_{\text{min}}(E)}{\tau}} \frac{1}{1 + \omega_0^2 \tau^2}.
\end{multline}
This term predicts a red-shift of the resonator frequency with increasing temperature. Notably, both the resonant- and relaxation-type couplings induce frequency shifts as a function of temperature, but are independent of power. While these expressions suggest that the scale of the TLS-induced frequency shift can be determined by the scale of the TLS-induced loss, this is only valid under the assumption of a uniform density of states. For the same reasons outline in Appendix \ref{appendix:relaxationTLSLoss}, we determine these parameters independently when fitting to the model.


In Fig. \ref{freqShift}(a-b), we plot the power-dependent frequency shift at $T_\textrm{bp}=10$ mK. The annealed \SiN{} devices shown in Fig. \ref{freqShift}(b) agree with the TLS model prediction that frequency is independent of drive strength. Curiously, we observe a strong power-induced frequency shift in devices with as-deposited \SiN{}. To understand this, we can look at the temperature-dependent frequency shifts at $\overline{n} \sim 1$ in Fig. \ref{freqShift}(c-d).  The temperature-dependent fractional frequency shift observed in the resonators with as-deposited or annealed \SiN{} is well described by fits that include the TLS couplings and a parameter-free contribution from thermal quasiparticles \cite{mattis_theory_1958}. Similar to the loss, we see that the resonators with annealed \SiN{} have negligible impact from the relaxation-type interaction compared to those with as-deposited \SiN{}. Noting the similarity in shape to the temperature-dependent frequency shift, we reuse the effective temperature $T_\text{eff}$ used to describe the loss in Fig. \ref{Q}. With this self-heating assumption, we find qualitative similarity between the data and the combined TLS and self-heating model. However, this combined model does not precisely predict the behavior of all of the resonators. This discrepancy likely stems from the previously mentioned fact that the loss and frequency shift expressions find support on different regions of the TLS density of states.

Another consequence of this frequency shift is that we must dynamically tune the frequency of the drive tone to maintain a 1.5 MHz detuning relative to the LC frequency. It is difficult to set the detuning appropriately for a given drive strength, because the drive detuning sets the intracavity photon number which may in turn shift the LC frequency. We converge on the correct detuning by iteratively setting the drive detuning relative to an assumed LC resonance frequency and then fitting the measured LC response. We repeat this loop until the predicted detuning differs from the measured detuning by less than 0.01\%.

\section{Spectroscopy Sample Preparation and Analysis}
\label{appendix:FTIR_details}
To prepare samples for the FT-IR, we deposit a 400 nm thick layer of \SiN{} onto both sides of a Si wafer. This film is much thicker than that used for the resonators because a thicker film will increase the absorption signal. We dice the wafer into four samples, and then etch the film on one side of each of the samples to obtain different thicknesses. This results in a set of samples with \SiN{} thickness ranging from 800 nm (no \SiN{} etched) to 500 nm (300 nm of \SiN{} etched on one side of the wafer). A similar procedure is used to prepare the annealed \SiN{} samples, except the anneal is performed directly after the deposition.

Our FT-IR spectrometer model is Thermo Scientific Nicolet iS50, operated in transmission mode \cite{boyle_quantitative_2022}. The transmission mode of operation is less sensitive than attenuated total reflection techniques, but nonetheless reveals the film impurities given proper sample preparation and data analysis. After purging the sample chamber with nitrogen gas for 30 minutes, we take a background spectrum with a bare Si wafer. For the spectra of each \SiN{} film, we subtract the background spectra and then perform an atmospheric correction to remove absorption features from residual water and CO$_2$ in the sample chamber. A polynomial fit is used to remove the baseline structure \cite{mazet_background_2005}. The areas under the absorption peaks are extracted from Gaussian fits to the spectra.

Using well-established calibrations of the infrared absorption cross-sections of these bonds, we can calculate the atomic hydrogen percentage in the film using the Lanford and Rand method \cite{lanford_hydrogen_1978}, described by
\begin{equation}
    \text{atomic \%H} = \frac{[\text{Si-H}] + [\text{N-H}]}{[\text{Si}] + [\text{N}] + [\text{Si-H}] + [\text{N-H}]},
    \label{percentHequation}
\end{equation}
where [Si], [N], etc. denote the densities in atoms per unit volume. We have made the assumption that all of the -H in the film are bonded to either N or Si. [Si-H] can be determined by $[\text{Si-H}] = A_{\text{Si-H}}/(\sigma_{\text{Si-H}} \times t)$, where $A_{\text{Si-H}}$ is the area under the Si-H absorption peak, $\sigma_{\text{Si-H}}$ is the IR absorption cross-section, and $t$ is the thickness of the film ([N-H] is found in an analogous manner). The sum [Si] + [N] is the atomic density of \SiN{} in units of atoms per unit volume, which can be determined from the density of the film.



\bibliography{references.bib}

\end{document}